\documentclass[twocolumn,twoside,slac_two]{revtex4}
\usepackage{graphicx}
\usepackage{fancyhdr}
\newcommand{\f}{4U\,1626--67 \,}
\newcommand{\fermigbm}{ {\it Fermi}/GBM }
\newcommand{\swiftbat}{{\it Swift}/BAT }

\newcommand{\be}{\begin{equation}}
\newcommand{\ee}{\end{equation}}
\newcommand{\bdm}{\begin{displaymath}}
\newcommand{\edm}{\end{displaymath}}

\begin{document}

\title{Discovery  of a  New  Torque  Reversal  of  the  Accreting  X-ray  Pulsar  \f  by  Fermi/GBM}

%

\author{A. Camero-Arranz}
\affiliation{Fundaci\'{o}n Espa\~{n}ola de Ciencia y Tecnolog\'{i}a (MICINN), C/Rosario Pino,14-16, 28020- Madrid, Spain}

\author{E. Beklen}
\affiliation{Middle East Technical University (METU), 06531 Ankara, Turkey}

\author{M. H. Finger}
\affiliation{Universities Space Research Association, 6767 Old Madison Pike,Huntsville, AL 35806}

\author{N.R. Ikhsanov, C.A. Wilson–-Hodge}
\affiliation{NASA/Marshall Space Flight Center, Huntsville, AL 35812}

\author{ P. Jenke}
\affiliation{ORAU/NPP, Oak Ridge, TN 37831}

\begin{abstract}

Recent X-ray observations  by Fermi/GBM  discovered a new torque reversal of \f  after 18 years of steady spinning down. Using Swift/BAT observations we were able to center this new torque reversal on Feb 4 2008, lasting approximately 150 days. From 2004 up to the end of 2007,  the spin–down rate averaged at a mean rate of  $\sim \dot{\nu}=-4.8 \times10^{-13}$\,Hz\,s$^{-1}$ until the torque reversal reported here. Since then it has been following a steady spin–up at a mean rate of    $\sim \dot{\nu}= 4 \times10^{-13}$\,Hz\,s$^{-1}$. The properties of this torque reversal, as well as the lack of correlation between the X-ray flux and the torque applied to the neutron star before this transition, challenges our understanding of the physical mechanisms operating in this system. 

\end{abstract}

\maketitle

\thispagestyle{fancy}


\section{INTRODUCTION}

 The accreting--powered pulsar \f was discovered
by {\it Uhuru} \cite{Giacconi72}. This low mass X--ray binary (LMXB)   consists of a 7.66 s X--ray pulsar accreting
from an extremely low mass companion (0.04 M$\odot$ for {\it i} = 18$^o$)
\cite{Levine88}. Although orbital motion has never been detected  in the  X--ray data, pulsed optical
emission reprocessed on the surface of the secondary revealed \cite{Middleditch81}
the 42 min orbital period.  The faint optical counterpart (KZ TrA, {\it V}$\sim$17.5)
has a strong UV excess and high optical pulse fraction \cite{McClintock77,McClintock80}.
A persistent 48 mHz quasi-periodic oscillation (QPO) has been detected in the X--ray emission
\cite{Shinoda90, Kommers98}. \cite{Orlandini98}  inferred a neutron star magnetic field in the range
(2.4--6.3)$\times$10$^{12}$ G. To compute this magnetic field
range a source distance of 5--13 kpc was assumed \cite[and references therein]{Chakrabarty98}.  A  $\sim$37 keV absorption cyclotron feature  was found in the 0.1--200 keV  {\it BeppoSAX} spectrum \cite{Orlandini98} . 

For more than a decade after the discovery of pulsations \cite{Rappaport77} the source underwent steady spin--up at a mean rate of $\sim \dot{\nu}=8.5 \times 10^{-13}$\,Hz\,s$^{-1}$ \cite{Chakrabarty97} (see Fig~\ref{pollas2}, Top). Monitoring of the source by the Burst and Transient Source Experiment (BATSE) on board the Compton Gamma Ray Observatory ({\it CGRO}) starting in  April 1991,  found the pulsar spinning down, implying a changed sign in the accretion torque \cite{Wilson93,Bildsten94}. During the 7 years after the first torque reversal, the pulsar  spun--down at a rate of $\sim \dot{\nu}=-7.2 \times 10^{-13}$\, Hz\, s$^{-1}$ \cite{Chakrabarty97}.

We present a long term timing and spectral analysis using all the available \fermigbm data since
its launch in 2008 June 11 and over 5 yr of hard X-ray \swiftbat data from2004 up to 2009. 

\section{Fermi/GBM}

\subsection{OBSERVATIONS}

Since  2008 June 11  \f   has been continuously  monitored by the
{\it Gamma-ray Burst Monitor} (GBM)\cite{Meegan09}, on board the
{\it Fermi} observatory. Timing analysis was carried out with GBM  CTIME data,
with  8 channel spectra every 0.256 seconds. The total exposure time was
$\sim$13.75 Ms. The  GBM is an all-sky instrument sensitive to X--rays and gamma rays with energies
between $\sim$8 keV and $\sim$40 MeV.  GBM includes 12 Sodium Iodide (NaI)
scintillation detectors and 2 Bismuth Germanate (BGO) scintillation
detectors. The NaI detectors cover the lower part of the energy range,
from 8  keV to about 1 MeV. The BGO detectors cover the energy range of
$\sim$150 keV to $\sim$40 MeV. Only data from the
NaI detectors were used in the analysis.

\subsection{TIMING ANALYSIS AND RESULTS}

\begin{figure}[!t]
\vspace{0.05cm}
\hspace{-0.625cm}\includegraphics[width=8.8cm,height=6.8cm]{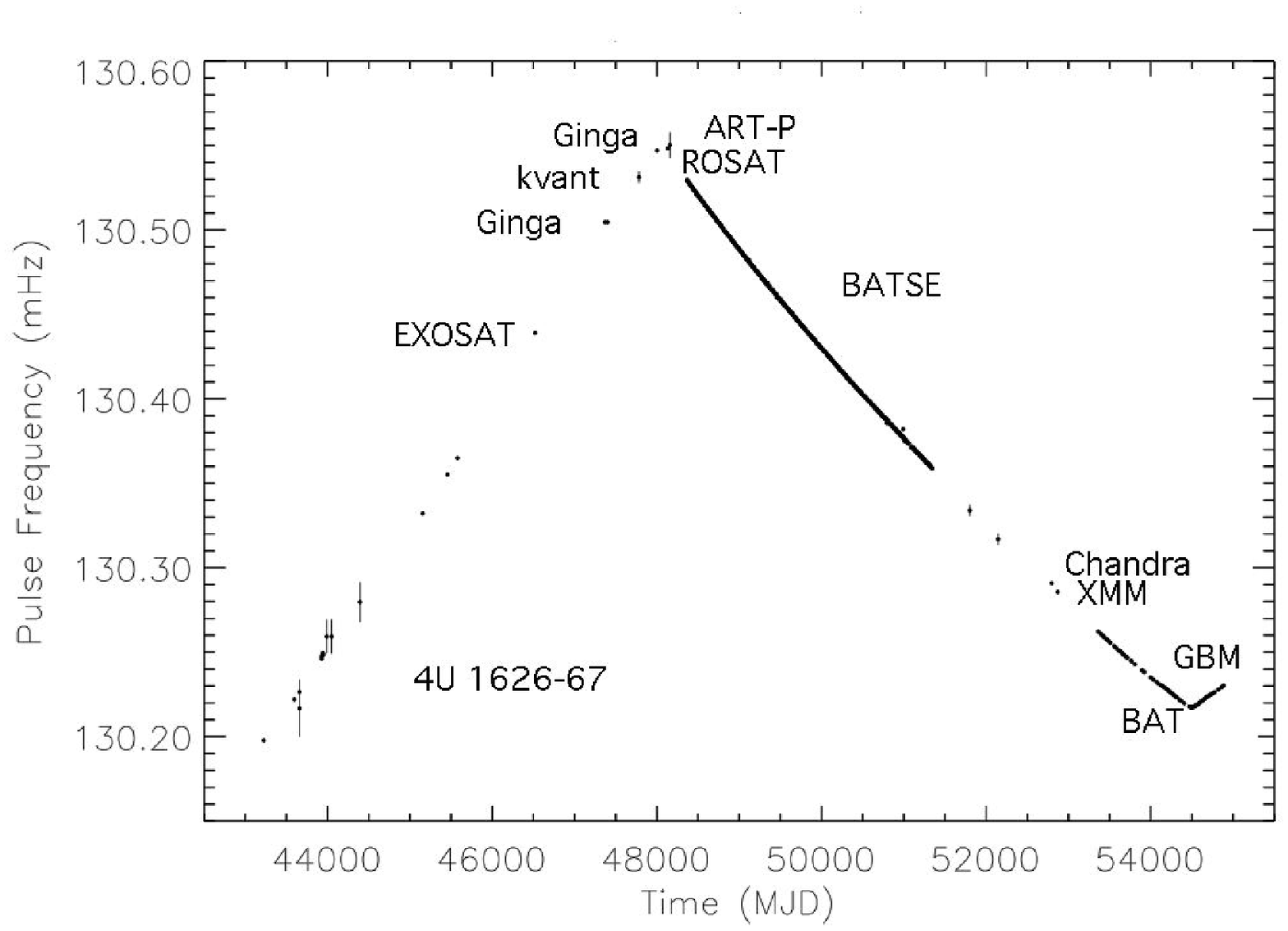}
\includegraphics[width=4cm,height=3.5cm]{Fig2a.eps}
\includegraphics[width=4cm,height=3.65cm]{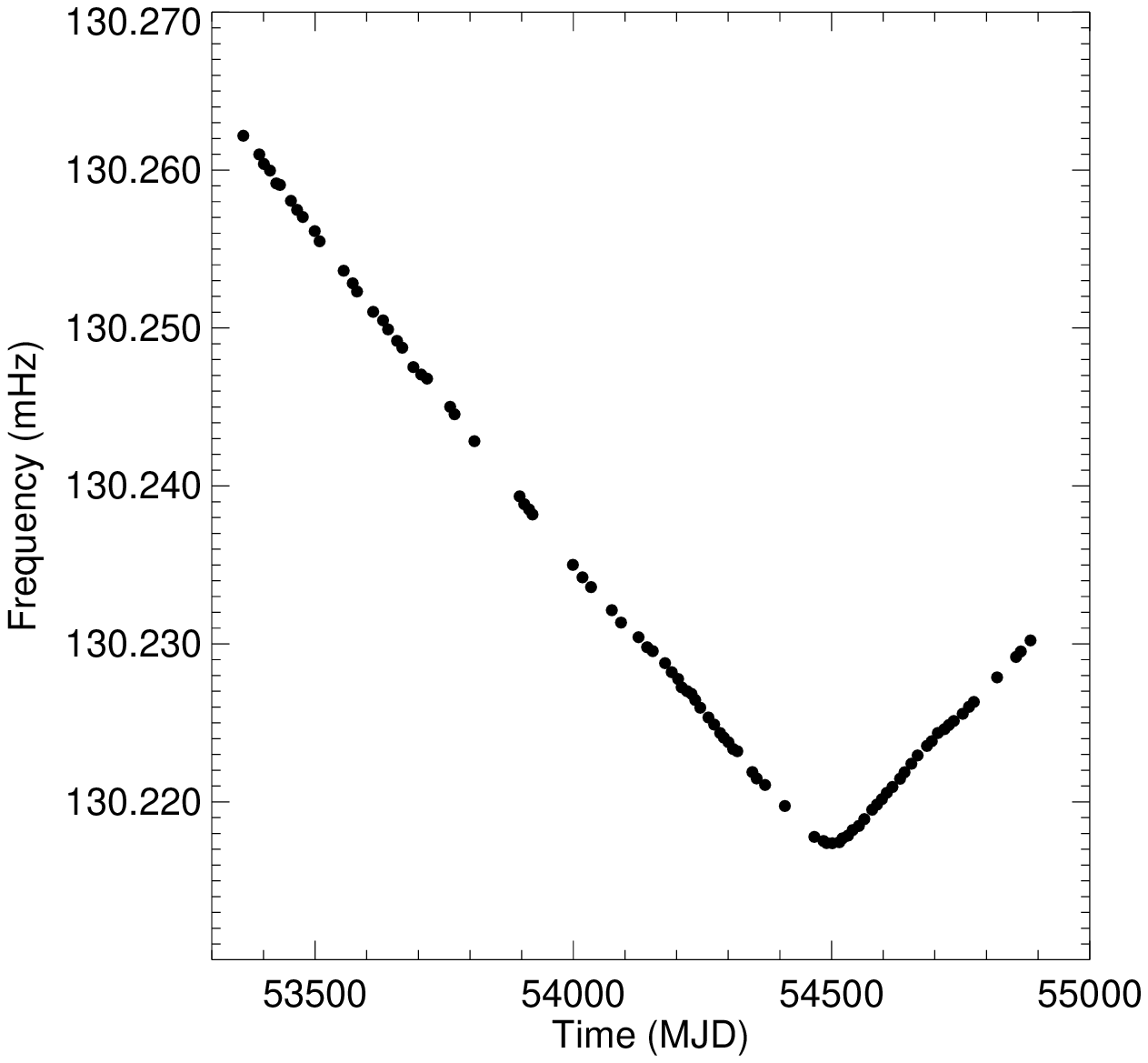}
 \caption{(Top). Pulse frequency history of 4U\,1626--67.  (Bottom Left). Zoom of the top panel, with Fermi GBM pulse frequency measurements since 2008 August. A change in the sign of the torque was found after 18 years of the source spinning down. (Bottom Right).  Zoom of the top panel, with \swiftbat pulse frequency history  covering the 2008 torque reversal. Error bars are smaller than the plotted symbols. \label{pollas2}}
\end{figure}
\vspace*{0.2cm}
All intervals of CTIME data from the 12 NaI detectors
are selected for analysis after excluding those containing
high voltage transients, phosphorescence events, rapid spacecraft slews,
South Atlantic Anomaly induced transients, electron precipitation events
and gamma-ray bursts (see Mark H. Finger et al., these proceedings).  Source pulses are then separated from
the background by fitting the rates in all detectors with a
background model, and subtracting the best fit model. Then we combine the residuals over detectors
with time dependent weights which are proportional
to the predicted (phase averaged) count rates from the pulsar.

Short intervals ($\sim$300s) of these combined residuals
are then fit with a constant plus a Fourier expansion to determine a
pulse profile. The profiles are divided into six day intervals and the pulse frequency and mean
profile determined in each interval with a search of pulse frequency for the maximum of the Y$_{n}$ (n=2)  statistic \cite{Finger99}. 

Our  monitoring of \f  with {\it Fermi}/GBM starting in 2008 August,  discovered  the pulsar spinning--up \cite{Camero09} . Fig~\ref{pollas2} (Right) shows the pulse frequency history using data from this monitoring. \f  seems to be increasing in $\dot{\nu}$. Follow-up {\it Fermi}/GBM observations  confirm that the pulsar it  is currently spinning--up  at a mean rate of $\sim\dot{\nu}$=4$\times10^{-13}$ Hz s$^{-1}$.

\section{Swift/BAT}

\subsection{OBSERVATIONS}

 The Swift Gamma-ray mission \cite{Gehrels04} was launched  on
2004 November 20. The hard X--ray (15--150 keV) Burst Alert Telescope
(BAT) on board {\it Swift},  monitors the entire sky and it produces continuous streams of {\it rate} data. 
 For the present study we  have analyzed  more than 4  years
of BAT {\it quadrant rates} observations (1.6 sec sampling; four energy bands; four separate spatial quadrants) 
 when  \f \,was visible (total exposure time $\sim13$\,Ms). For the hardness ratio analysis
we used count rates from the \swiftbat transient monitor results
provided by the \swiftbat  team\footnote{http://heasarc.gsfc.nasa.gov/docs/swift/results/transients}.

\subsection{ TIMING ANALYSIS  AND  RESULTS}


\begin{figure}
\vspace{0.4cm}
 \includegraphics[width=7.5cm,height=8cm]{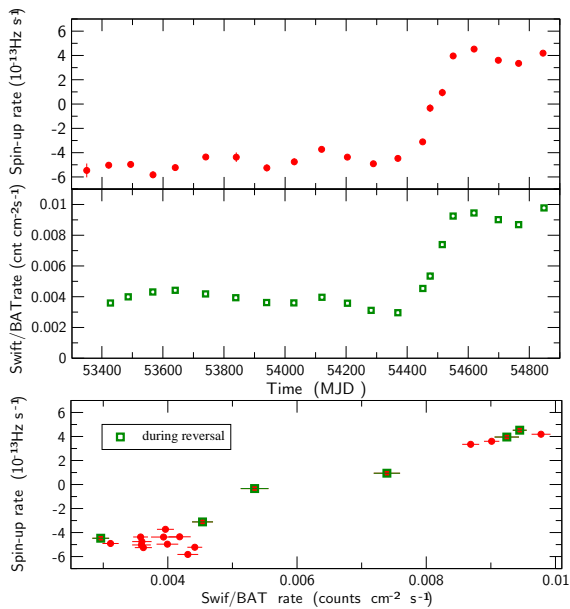}
 \caption{Top panel. \swiftbat spin--up rate history of \f. Middle panel. Average 15--50 keV BAT count rate vs. time. Error bars are  smaller than the plotted symbols. Bottom panel. BAT count rate vs. spin--up rate for all the period (circles). A correlation pattern is observed specially during the torque reversal (only square symbols).  \label{fdot}}
\vspace{0.3cm}
\end{figure}


A similar procedure was followed for the \swiftbat {\it quadrant rates}
timing analysis. Initial good time interval (GTI) files are obtained using the {\it maketime}
ftool (heasoft-6.6.1)\footnote{http://heasarc.gsfc.nasa.gov/docs/software.html}. Then a filtered version  of the {\it quadrant rates} is obtained,  to then finally be  barycentered using the ftool {\it barycorr}.
 With the ftool  {\it batmasktgimg} the pixel exposure fraction for each quadrant is
computed for the center of  each (refined)   GTI interval.  Pulse
profiles for  each good  GTI interval are computed. First the rates
for each quadrant  are fit to a  quadratic+Fourier expansion. Then
the Fourier coefficients are combined using the quadrant  exposures
 to produce mean profiles (with units of  counts s$^{-1}$ cm$^{-2}$). In a
final stage,  the  Y$_{n}$ (n=2) statistic is again used in intervals of 35
days and a frequency search for pulsations is carried out.  The spin rates were
computed by  fitting a linear function to the frequencies, which were divided into 21 time intervals.

\swiftbat observations allowed us to cover  the evolution of this
second torque reversal. We found that the pulsar
spun--down  at a mean rate of $\sim \dot{\nu}=-4.8 \times 10^{-13}$\,Hz\,s$^{-1}$ until the source reversed
torque. Fig~\ref{pollas2}  (Left) shows that the transition took place at
around MJD 54500  (2008 Feb 04) and lasted  approximately  150 days.   In the bottom
panel of Fig~\ref{fdot} we can see that  there is a strong correlation between the
\swiftbat count rate and the spin--up rate especially  during the reversal.  We have not observed any significant change in pulse shape, not even during the reversal.

\section{RXTE}

\subsection{OBSERVATIONS}

The Rossi X--ray Timing Explorer ({\it RXTE}) \cite{Bradt93} carries 3 instruments on board. The Proportional
Counter Array (PCA) \cite{Jahoda96} (2--60 keV), the High Energy X-ray Timing Experiment (HEXTE) \cite{Gruber96} (up to 200 keV) and   the All Sky Monitor (ASM) \cite{Levine96}  (2--10 keV). Two {\it RXTE}/PCA observations from 2008 March 5 and 13 were used (ID 93431--01--01--00 and 93431--01--02--00; 7.174 ksec).  For spectral analysis
we selected PCA Standard--2 data  and  HEXTE Standard Modes (Archive) data (64-bin spectra every 16s). For the long--term hardness  ratio analysis we used  the ASM daily flux averages in the 1.5--12 keV energy range from the HEASARC  archive\footnote{http://heasarc.gsfc.nasa.gov/docs/archive.html}.

\subsection{SPECTRAL ANALYSIS AND RESULTS}

{\it RXTE}/PCA (2.5--20 keV) and  HEXTE (18--100 keV) spectra were fitted in XSPEC 11.3.2  with two models   used by \cite{Pravdo79}.  Using these models  allows us to compare our spectral study  
with previous works  by \cite{Pravdo79,Orlandini98,Krauss07,Jain09} and update the  long-term  X--ray flux history of 
4U\,1626--67  relative to the flux measured by {\it HEAO 1} (Chakrabarty et al. (1997); Krauss et al. (2007))  .
The first  model  includes a low-energy absorption, a blackbody component, a power law and a high-energy  cutoff at 
$\sim$20 keV  (WABS (GAUSS+BBODY+POWLAW) HIGHECUT).  A broad line near 6.5 keV  significantly improves the present  fit and  indicates the presence of an iron line, also suggested by \cite{Pravdo79} in their ( 0.7--100 keV) spectral  analysis of this source. The column density of cool material in the line of sight was fixed in our study since it could  not be constrained. A value of $1.3 \times 10^{21}$\,cm$^{-2}$ was selected from \cite{Krauss07}.
The spectral parameters obtained are shown in Table 1.  We fit in addition the same model with a bremsstrahlung instead of a blackbody component , obtaining  a  compatible fit. Table~\ref{specfits}  summarizes the  spectral  
parameters obtained.


\begin{table*}
  \caption{ {\it RXTE}/PCA and HEXTE CONTINUUM SPECTRAL FITS$^{mod}$ }
  \begin{tabular}{lllllllllll}
  \hline\noalign{\smallskip}

  \hline\noalign{\smallskip}
  Observation   & $\alpha^*$  & E$_{cut}^{\dag}$       & E$_{Fold}^{\dag}$           & Gaussian$^{\dag}$   & Gauss.
                     & Gauss.         & T$_{BBody}^{\dag}$    &  norm  & Flux$^{***}$         & $\chi_r^{2}$(DOF)  \\
       (MJD)     &         &    &    &     &  $\sigma^{\dag}$
                  & norm$^{**}$   & T$_{Bremss}^{\dag}$        &   $^{BBod^1}_{Brem^2}$       &            &  \\
  \hline\noalign{\smallskip} \hline\noalign{\smallskip}

93431--01--01--00   & 0.75(2) & $18.2{+0.1\atop -0.3}$   & 8.5(4)  & $6.05{+0.5\atop -0.17}$ & $1.6{+0.3\atop -0.2}$ & $2.4{+4\atop -0.8}$     & $0.615{+0.006\atop -0.018}$ &    0.0013      & 1.01(8) & 1.15(114)      \\

    (54530)      & $0.74(2)$& $18.2{+0.3\atop -0.2}$ & $8.5{+0.3\atop -0.4}$ & $6.42{+0.17\atop -0.4}$ & $1.4{+0.1\atop -0.2}$ & $1.5{+1.2\atop -0.6}$ & $1.74{+0.12\atop -0.14}$ & 0.108   & 1.01(1) & 1.15(114)   \\

  \hline\noalign{\smallskip}

 93431--01--02--00  & $0.71{+0.06\atop -0.04}$ & $17.90{+0.19\atop -0.3}$  & $8.4{+0.5\atop -0.6}$    & $6.2{+0.6\atop -0.4}$ & $1.5{+0.4\atop -0.2}$ & $2.23{+1.9\atop -0.08}$  & $0.654{+0.04\atop -0.009}$      &   0.0013      & 1.006(12) & 1.29(114)\\

   (54538)      & $0.71{+0.04\atop -0.05}$  & $17.96{+0.14\atop -0.2}$  & 8.4(6)  & 6.9(2) & $0.5{+0.4\atop -0.3}$  & $0.396{+0.14\atop -0.015}$ & $2.47{+0.3\atop -0.08}$  &  0.068    & 1.004(9)  &    1.28(114)\\

   \hline \noalign{\smallskip}

   \end{tabular}
    
\hspace{-0.38cm} $^{mod}$WABS (GAUSSIAN+ BLACKBODY /  BREMSSTRAHLUNG +POW) HIGHECUT  ($N_{\rm H} = 1.3 \times 10^{21}$\,cm$^{-2}$   fixed, \\
 \hspace{-13cm}   uncertainties  at 3$\sigma$ level) \\
  \hspace{-14.9cm}    $^*$ Photon Index \\
   \hspace{-16.3cm}   $^{\dag}$  keV \\
    \hspace{-15.9cm}  $^{**}$    $\times 10^{-3}$ \\
     \hspace{-12.2cm}   $^1$  (Lumin/$10^{39}$\,erg\,s$^{-1}$)(d/$10kpc)^{-2}$\\
    \hspace{-11cm}  $^2$  $3.05\times 10^{-15} (4 \pi d^2)^{-1} \times$Emission measure  \\
 \hspace{-12cm}      $^{***}$  $\times 10^{-9}$\,erg\,cm$^{-2}$\,s$^{-1}$    (2--100 keV)
\label{specfits}
\end{table*}


\section{HARDNESS RATIO ANALYSIS}

\begin{figure}
 \includegraphics[width=7.75cm,height=4.9cm]{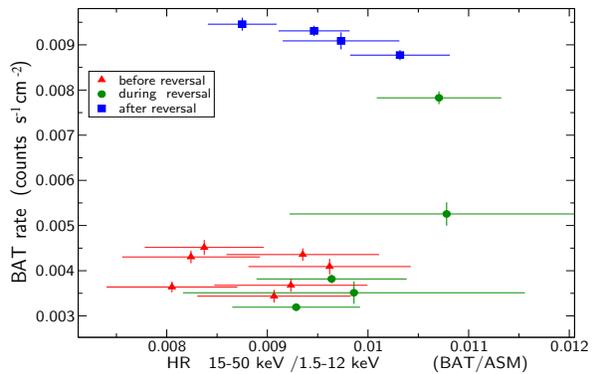}
\caption{Hardness--intensity diagram of 4U\,1626--67.  For HR,   \swiftbat count rates (15--50 keV) were selected as the hard band  and {\it RXTE}/ASM count rate (1.5--12 keV) as the soft band. During the reversal a transition from hard to soft is seen (green circles). \label{hr}}
\vspace{0.4cm}
\end{figure}


Fig~\ref{hr}  shows Hardness--intensity diagram (HID) for \f using  BAT count rate.
The hardness ratio (HR)  was defined  as the ratio 15--50keV/1.5--12keV
(BAT/ASM). To reduce large uncertainties the light curves were rebinned and
 then the HR were computed.   This allow us to study  the  long-term 
spectral variability of  \f,   including  the transition,  since the 2 RXTE  observations do not 
provide us  any direct comparison between before and after the torque reversal. From that figure 
we can see that there is a transition from hard  to soft during this new reversal of 4U\,1626-67.

\section{DISCUSSION}

All previous studies of \f were focused on modeling the spin-up torque
applied to the neutron star from the accreted material. It was widely believed that the
spin behavior of the pulsar depended mainly on variations of the mass accretion rate onto
the stellar surface and therefore, the rate of mass transfer between the system
components.  A dramatic change in the power spectra between the last observation of \f during the spin-down phase (2003)  \cite{Kaur08}, and observations made soon after the new torque reversal has been recently reported by \cite{Jain09}, with the 35--48 mHz QPO no longer being present, and wide shoulders on the pulse fundamental appearing. They claimed that the observed behavior of the source cannot be a simple case of increased mass transfer rate, but is also a change in the accretion flow parameters.

Analyzing the evolution of the source energy spectrum and possible correlation between
the torque and X-ray luminosity of the pulsar,  \cite{Yi99} proposed a scenario in which
the torque reversal in 1990 is associated with a state transition of the accretion disk to a
geometrically thick, hot and, possibly, sub-Keplerian phase. Following this idea one
could associate the 2008 torque reversal with an inverse transition of the disk into its
previous geometrically thin Keplerian phase. However, the reason for such a transition
is rather unclear since the level of X-ray flux measured before and even after the 2008
reversal is smaller than that measured during the reversal in 1990. Furthermore, both
reversals have occurred at almost the same timescale (about 150\,days), which
significantly exceeds the dynamical timescale in the hot disk in which its transition to
the ground state is expected.  The difficulty to fit  the transition timescales observed
in \f has also been mentioned by \cite{Wijers99}, who discussed a
possibility to explain the torque reversals in terms of the warped disk transition into
a retrograde regime.

\begin{figure}
\includegraphics[width=7.75cm,height=4.9cm]{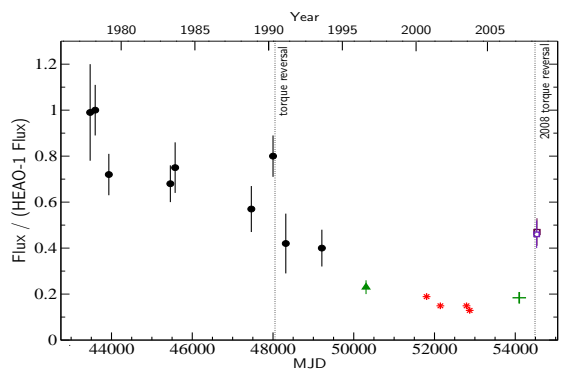}
 \caption{The X--ray flux history of 4U\,1626--67  relative to the flux measured by {\it HEAO 1},
in the same energy band, from previous works (Chakrabarty et al. (1997): circles; Orlandini et al.  (1998): triangle; Krauss et al. (2007): stars) and two recent {\it RXTE}/PCA observations (unfilled squares) in the 2--20 keV band. The cross point is inferred from PCA flux and the fractional change in the \swiftbat rate.\label{fluxhist}}

\end{figure}


A correlation between the torque applied to the neutron star in \f and X-ray flux of the
system in the above mentioned models has been adopted as one of the basic assumptions.
To test the validity of this assumption using data derived before 1993 was rather
complicated. This is illustrated in Figure~\ref{fluxhist}, which shows the \f X-ray flux
history. We can see all previous flux measurements and two {\it RXTE}/PCA recent values
from the present work (in the 2--20\,keV band). These values are relative to the flux
measured in 1978 by {\it HEAO 1} in the same energy band \cite[][and references
therein]{Orlandini98, Krauss07, Chakrabarty97}. The cross point before the 2008 reversal
has been inferred by scaling the PCA fluxes according to the observed change (2.5
factor) in the \swiftbat rate, since  no spectral changes across the transition have
been observed according to the present work.

As seen from Figure~\ref{fluxhist}, the X-ray flux during the spin-down phase has
decreased by a factor of 2. This indicates that the mass accretion rate onto the surface
of the neutron star, $\dot{M}$, and, correspondingly, the spin-up torque applied to the
star \cite{Pringle72},
 \be
K_{\rm su} = \dot{M} (GM_{\rm ns} r_{\rm m})^{1/2},
 \ee
during this phase have also decreased by at least the same value. If the spin-down
torque applied to the neutron star during this time were constant one would expect the
pulsar to brake harder at its fainter state close to the end of the spin-down phase.
However, observations show the situation to be just the opposite. The spin-down rate of
the neutron star during this phase has decreased from $|\dot{\nu}| \simeq 7 \times
10^{-13}\,{\rm Hz\,s^{-1}}$ \cite{Chakrabarty97} to $5 \times 10^{-13}\,{\rm
Hz\,s^{-1}}$ (see Figure~\ref{pollas2}), implying that the pulsar was braking harder at
its brighter stage just after the torque reversal in 1990. According to the equation
governing spin evolution of an accreting neutron star,
 \be
2 \pi I \dot{\nu} = K_{\rm su} - K_{\rm sd},
 \ee
this means that the spin-down torque, $K_{\rm sd}$, during the spin-down phase has been
decreasing simultaneously with the spin-up torque but at a higher rate and, therefore,
the pulsar spin evolution during this time has been governed mainly by variations of
$K_{\rm sd}$ rather than $K_{\rm su}$ (here $I$ is the moment of inertia of the neutron
star). This conclusion seriously challenges the possibility of modeling the spin history
of \f solely in terms of variations of $\dot{M}$, and suggests that  the dramatic
increase of X-ray flux observed in 2008 torque reversal may be a consequence rather than
a reason for this event.

With the  lack of correlation between the X-ray flux and the torque applied to the
neutron star,  modeling of the spin-down torque appears to be the main target for
theoretical studies of the system. Unfortunately, this part of modeling of the
magneto-rotational evolution of neutron stars remains  so far a work in progress.

Perna et al. (2006)  proposed a model where simultaneous accretion from a disk onto the neutron star,
some material from near the disk-magnetosphere boundary is ejected,  and either escapes from the system or is
recycled back into the accretion disk.  Their model predicts, however, that
the luminosity after a spin-down to spin-up torque reversal would be higher than the
luminosity after a spin-up to spin-down torque reversal, which is the opposite of what
occurred for 4U~1626-67 for this new reversal. Moreover, for 4U~1626-67 they predicted
the full spin-down/spin-up cycle would take thousands of years, again inconsistent with
the present observations.

Finally, the spectral evolution of 4U 1626-67 during the torque reversal differs from
 that expected in models which suggest significant changes of the accretion flow
structure in spin-up/spin-down transitions  [e.g.]\cite[]{Yi99,
Wijers99}. As seen from Figure~\ref{hr}, the spectrum becomes the hardest during
the reversal and the value of the hardness ratio before and after these events does not
differ significantly. This indicates that the recent torque reversal can be associated
with changes of physical conditions in the inner part of the disk or/and in the
region of its interaction with the magnetosphere rather than a significant change of the
accretion flow geometry. The errors of the observations are, however, too large for a
justification of particular transition model.

\section{CONCLUSIONS}

We report on a discovery of a new spin-down to spin-up torque reversal in 4U~1626-67. It
occurred after about 18\,years of the pulsar's steadily spinning down and was centered
on 2008 Feb 4. The transitions lasted $\sim$150 days and was accompanied by an increase
in the {\it Swift}/BAT count rate of a 2.5 factor ($\sim$150$\%$). The pulsar spectrum
was harder during the torque transition than before or after. A strong correlation
between torque and luminosity is inferred only during the transition. The spin-up and
spin-down rates before and after the transition were almost identical ($\sim
\mid\dot{\nu}\mid=5 \times10^{-13}$\,Hz\,s$^{-1}$). However, the pulsar was braking
harder at the beginning of the spin-down epoch in 1990 than at its end in 2008.
Furthermore, the spin-down rate during this epoch was decreasing simultaneously with the
decreasing of the source X-ray luminosity. Finally, the spin-down to spin-up torque
reversal in 2008 occurred at lower luminosity as the spin-up to spin-down torque in
1990. These properties cannot be explained with existing models and  appear to be a clue 
for further progress in understanding the mechanism governing the torque reversals in the 
accretion-powered pulsars.

\begin{acknowledgments}
A.C.A. thanks for the support of this project to the Spanish Ministerio de Ciencia e Innovaci\'on through 
the 2008 postdoctoral program MICINN/Fulbright under grant 2008-0116. N.R.I. acknowledges supported from NASA
Postdoctoral Program at NASA Marshall Space Flight Center, administered by Oak
Ridge Associated Universities through a contract with NASA, and support from Russian
Foundation of Basic Research under the grant 07-02-00535a. M.H.F. acknowledges support
from NASA grant NNX08AG12G.  We also want to thank all the {\it Fermi}/GBM team for its help.

\end{acknowledgments}

\bigskip 

\end{document}